\documentstyle[a4,epsf]{article}
\begin{document}
{\Large\bf Soft pions at high energy and the flavor asymmetry
of the light sea quarks in the nucleon}\\
\begin{center}
Susumu {\sc Koretune}\footnote{E-mail address: koretune@shimane-med.ac.jp}\\
Department of Physics,Shimane Medical University,\\
Izumo,Shimane,693-8501,Japan\\
\end{center}

The modified Gottfried sum rule makes clear
importance of the high energy region not only in the theoretical
meaning but also in the numerical analysis.
In this talk, it is shown that
the soft pion theorem in the inclusive reaction at high energy can
explain the magnitude about $0.02 \sim 0.04$ in the NMC deficit.
The main contribution comes from the small $x$ region.
We also estimate the soft pion contribution to the Ellis-Jaffe sum rule and
show it to be negligible. However we find that the contribution to $g_1^{ep}$
becomes positive below $x = 0.002$.\\
\section{Introduction}
The modified Gottfried sum rule \cite{Got} has explained the NMC deficit in the
Gottfried sum \cite{NMC} almost model independently. It has shown that
the deficit is the reflection of the hadronic vacuum originating from the
spontaneous chiral symmetry breaking. In this sense the physics underlining
this algebraic approach has a common feature with that of the mesonic 
models reviewed in Ref.\cite{meso}. However,in the algebraic approach,
importance of the high energy region not only in the theoretical
meaning but also in the numerical analysis has been made clear.
Further the numerical prediction based on this
sum rule exactly agrees with the recent experimental value from E866/NuSea
collaboration \cite{E866}. This experiment also gives us the light antiquark
difference $(\bar{d}(x)-\bar{u}(x))$ and the ratio $\bar{d}(x)/\bar{u}(x)$
in the range $0.02\leq x \leq 0.345$. An unexpected behavior is that the
asymmetry seems to dissapear at large $x$. On the other hand, a typical
calculation in the mesonic models based on the $\pi NN$ and the $\pi N\Delta$ 
processes account for about a half of the NMC deficit \cite{meso}. 
According to the E866 experiment, an explanation of the remaining half of 
the NMC deficit should be given by contributions in  
the small $x$ region. Unfortunately, the approach from the mesonic models
can not account for the magnitude from these regions definitely.
In fact, the $\pi N\Delta$ process partly cancels the positive contribution
to the $(\bar{d}(x)-\bar{u}(x))$ from the $\pi NN$ process. The contributions
from the higher resonances or from the multiparticle states are obscure.
Hence the best we can say is that the mesonic models explain the flavor
asymmetry of the light sea quarks qualitatively. These facts suggest
that there may exist a dynamical mechanism so far overlooked to produce
the flavor asymmetry at medium and high energy, and that it may compensate
the above flaw of the mesonic models. In this talk, it is shown that
the soft pion theorem in the inclusive reaction at high energy \cite{sakai}can
explain the magnitude about $0.02 \sim 0.04$ in the NMC deficit.  The theoretical
prediction depends heavily on the spin dependent structure function $(g_1^{ep}-g_1^{en})$,
and the main contribution to the NMC deficit in this theorem comes from the small $x$ region.
Thus, this type of the contribution may become the background of the mesonic model.
\section{Soft pion at high energy}
Here we briefly explain the soft pion theorem in the inclusive reactions.
Usually,the soft pion theorem has been considered to be applicable only in the low
energy regions. However in Ref.\cite{sakai}, it has been found that this theorem
can be used in the inclusive reactions at high energy if the Feynman's
scaling hypothesis holds. In the inclusive reaction ``$\pi + p \to
\pi_{s}(k) + anything$'' with the $\pi_{s}$ being the soft pion, 
it states that the differential cross-section
in the center of the mass (CM) frame defined as
\begin{equation}
f(k^3,\vec{k}^{\bot},p^0)=k^0\frac{d\sigma}{d^3k} ,
\end{equation}
where $p^0$ is the CM frame energy, scales as
\begin{equation}
f\sim f^{F}(\frac{k^3}{p^0},\vec{k}^{\bot}) + \frac{g(k^3,\vec{k}^{\bot})}{p^0} .
\end{equation} 
If $g(k^3,\vec{k}^{\bot})$ is not singular at $k^3=0$, we obtain
\begin{equation}
\lim_{p^0\to \infty}f^F(\frac{k^3}{p^0},\vec{k}^{\bot}=0)=
f^F(0,0)=\lim_{p^0\to \infty}f(0,0,p^0) .
\end{equation}
This means that the $\pi$ mesons with the momenta $k^3<O(p^0)$ and
$\vec{k}^{\bot}=0$ in the CM frame can be interpreted as the soft pion.
This fact holds even when the scaling violation effect exists,
since we can replace the exact scaling by the approximate one in this
discussion.  The important point of this soft pion 
theorem is that the soft-pion limit can not be interchanged with the
manipulation to obtain the 
discontinuity of the reaction ``$a + b + \bar{\pi_s} \to a + b + \bar{\pi_s}$''. 
We must first take the soft pion limit in the reaction ``$a + b \to \pi_s + anything$''.
This is because the soft pion attached to the nucleon (anti-nucleon)
in the final state is missed in
the discontinuity of the soft pion limit of
the reaction ``$a + b + \bar{\pi_s} \to a + b + \bar{\pi_s}$'' \cite{sakai}.
\section{Kinematics}
Now we consider the soft pion theorem in the semi-inclusive currents-nucleon
reaction with use of the light cone variable \cite{kore78}. We take the
electromagnetic current for an illustration. We first take 
$k^{+}=0$ and $\vec{k}^{\bot}=0$, and then take $k^{-}=0$. In this limit,
$k^2=0$ but the momentum of the virtual $\gamma$ and the nucleon are unrestricted.
The amplitude in this limit is classified into three terms as is shown
in Fig.1. The graph (a) is the
contribution from the pion emission from the initial nucleon,the graph (b) is the
one from the final nucleon or the anti-nucleon, and the graph (c) is the term
coming from the null-plane commutation relation at $x^{+}=0$. 
\begin{figure}
   \epsfxsize = 15 cm   %or \epsfysize = 10 cm
   \centerline{\epsfbox{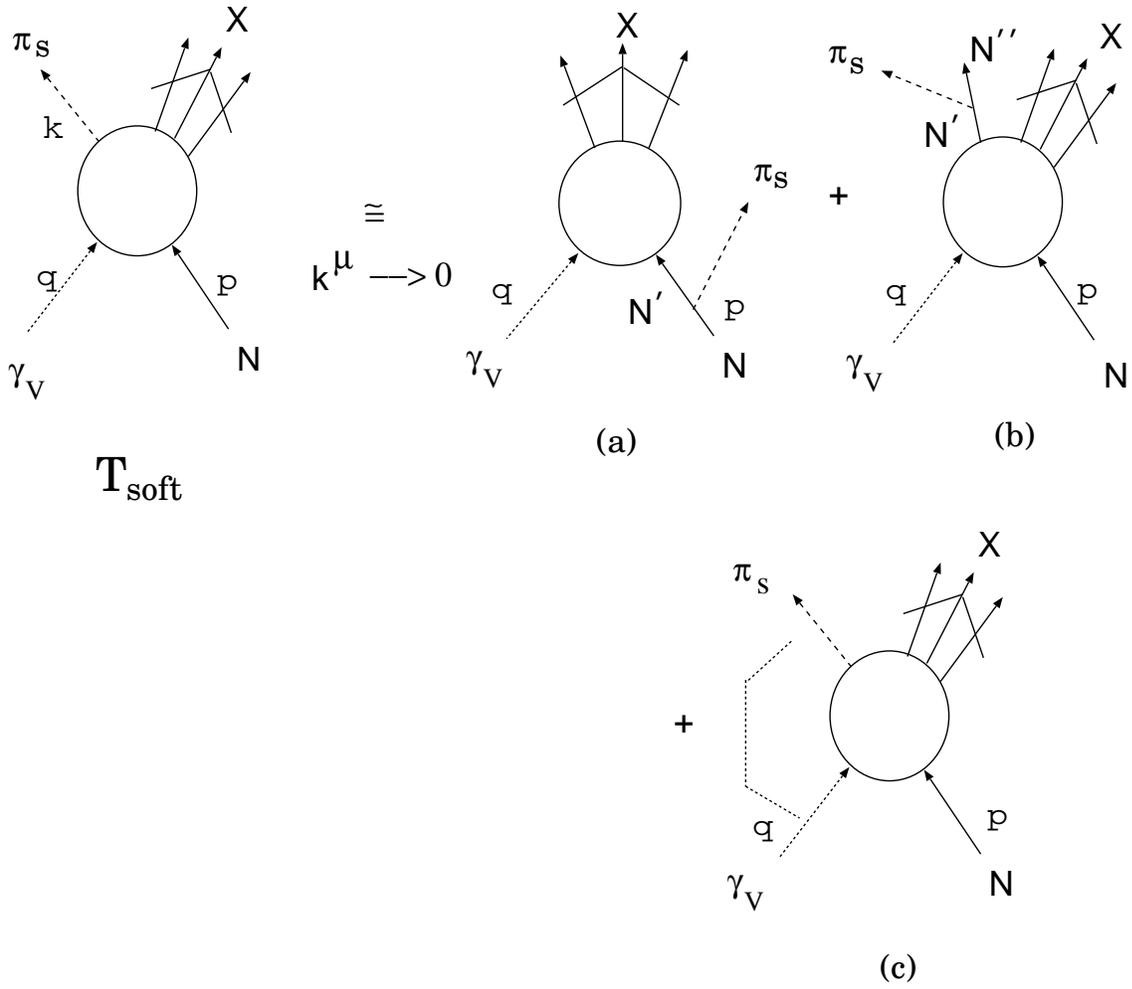}}
   \caption{Soft pion theorem in the inclusive reaction. the graph (a) corresponds
to the pion emission from the initial nucleon, the graph (b)corresponds to the final nucleon or 
the anti-nucleon, and the graph
(c) corresponds to the term coming from the null-plane commutator.}
        \label{fig:1}
\end{figure}
Then the hadronic tensor can be obtained by squaring the amplitude and the typical
graphs contributing to this tensor are shown in Fig.2.
\begin{figure}
    \epsfxsize = 15 cm   %or \epsfysize = 10 cm
    \centerline{\epsfbox{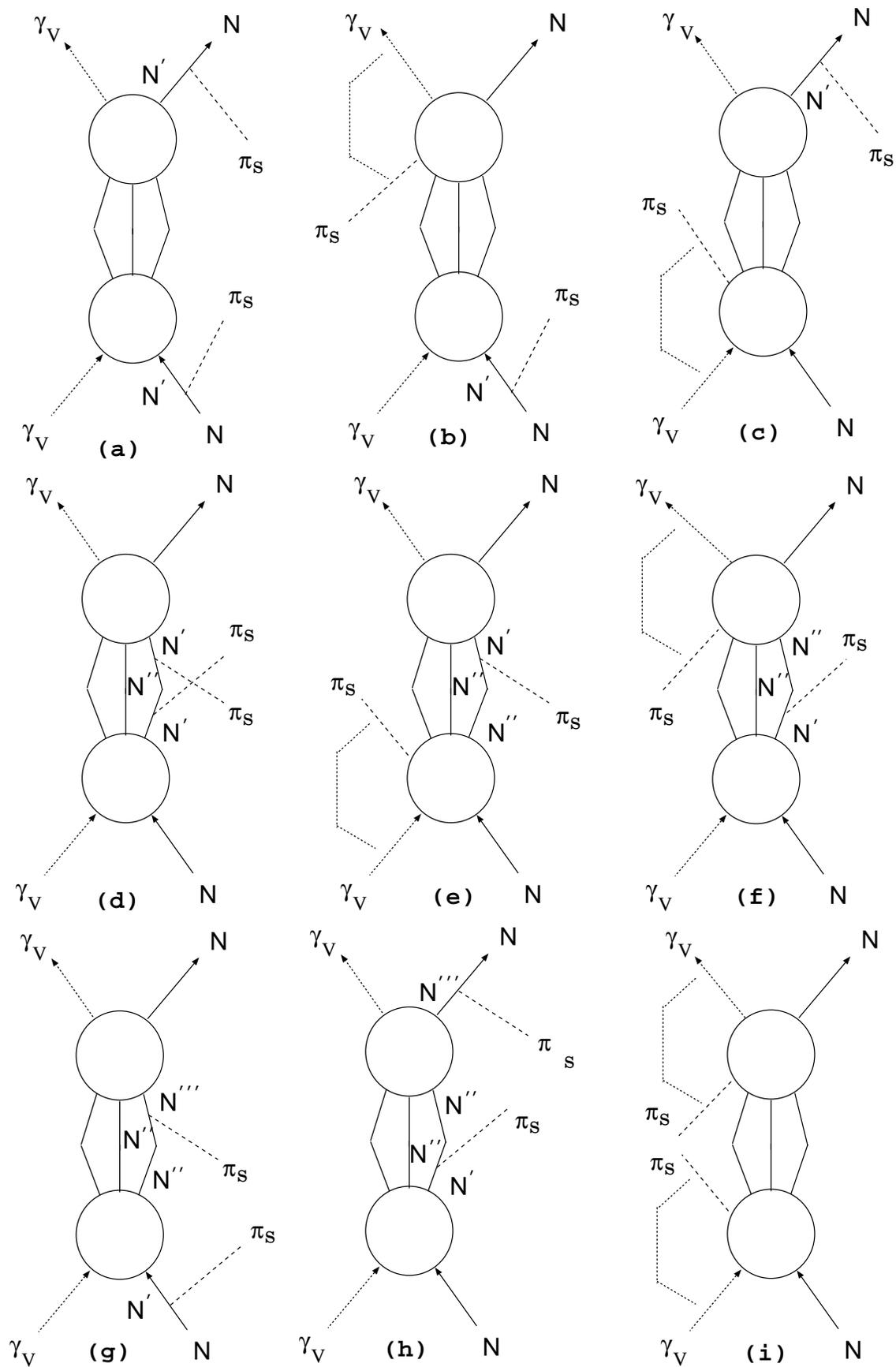}}
    \caption{The graphs contributing to the hadronic tensor}
    \label{fig:2}
\end{figure}
We discard the graphs (e)$\sim$(h) in Fig.2. These graphs are
characterized by the one soft pion from the nucleon (anti-nucleon) in the final state.
Because this emission vertex is  proportional to the helicity of the
nucleon (anti-nucleon),
the positive nucleon (anti-nucleon) and the negative one in the final state cancels each other 
at high energy, while at low energy these graphs are suppressed in the deep inelastic region
by the form factor effects. Among the remaining graphs, the graphs (a) and (d)
are directly related to the known processes. However the graphs (b),(c),and (i)
can not in general be related to the known processes without further assumptions.
Now in the deep inelastic region, these graphs are light-cone dominated, hence
we can use the perturbative QCD such as the cut vertex formalism\cite{Mue}. To obtain
the relation to the structure functions in the total inclusive reactions, however, we must
use the inverse transform of the moment sum rules. The net result of such
an analysis can be obtained by using the light-cone current algebra \cite{fg} at some initial
$Q^2=Q_0^2$ where the evolution is started and then to take 
the $Q^2$ dependence into account through the structure functions related by this way.
\section{The charge asymmetry}
The method in Ref.\cite{sakai} had not been checked experimentally, hence it was
done in the soft $\pi^{-}$ case \cite{kore78reso}. From the experimental data
of the Harvard-Cornell group\cite{harvard} the data satisfying the following conditions
\begin{enumerate}
\item[(1)]The transverse momentum satisfys $|\vec{k}^{\bot}|^2\leq m_{\pi}^2$.
\item[(2)]The change of $F^-$ can be regarded to be small in the small $x_F$ region.
\end{enumerate}
are selected. The effective cut of $x_F$ is about $0.2$. Then the theoretical value 
is about 10\% $\sim$ 20\% of the experimental value. However,
in the central region, there are many pions from the decay of the resonances, and about
20\% $\sim$ 30\% can be expected to be the pion from the directly produced pion.
Hence the theoretical value is the same order with the experimental value.
Now to reduce the ambiguity due to the pion from the resonance decay product,
the charge asymmetry was calculated\cite{kore82}. By assuming symmetric sea polarization
for simplicity, we obtain
\begin{eqnarray}
F_2^+ - F_2^- & =\frac{1}{4f_{\pi}^2}[F_2^{\bar{\nu}p}-F_2^{\nu p} +2g_A^2(0)F_2^{en}
-8xg_A(0)(g_1^{ep}-g_1^{en})\nonumber \\
& + 2g_A^2(0)(<n>_p + <n>_{\bar{n}}-<n>_{\bar{p}}-<n>_n)F_2^{ep}],
\end{eqnarray}
\begin{eqnarray}
F^+ - F^- & = \frac{1}{64\pi^3f_{\pi}^2}[\frac{F_2^{\bar{\nu}p}-F_2^{\nu p}}{F_2^{ep}}
+2g_A^2(0)\frac{F_2^{en}}{F_2^{ep}}-8xg_A(0)\frac{(g_1^{ep}-g_1^{en})}{F_2^{ep}} \nonumber \\
& + 2g_A^2(0)(<n>_p + <n>_{\bar{n}}-<n>_{\bar{p}}-<n>_n)].
\end{eqnarray}
where $F^{\alpha}$ is defined as
\begin{equation}
F^{\alpha}=\frac{1}{\sigma_T}q^0\frac{d\sigma^{\alpha}}{d^3q}.
\end{equation}
By neglecting the nucleon multiplicity term which can be expected to be a small positive
contribution, the theoretical value is roughly equal to $0.15 \sim 0.18$. While the
experimental value with the transverse momentum satisfying
 $|\vec{k}^{\bot}|^2\leq m_{\pi}^2$ in Ref.\cite{harvard} 
is almost constant in the region $ 0< x_F < 0.1$ with its value $0.28 \pm 0.05$,
and it gradually decreased above $x_F = 0.1$.
Hence the theoretical value is very near to the experimental value.
\section{The soft pion contribution to the Gottfried sum}
The charge asymmetry in the central region may contribute to
the Gootfried sum and hence its magnitude has been estimated \cite{kore99}.
Adding the contributions from the soft $\pi_s^+,
\pi_s^-,$ and $\pi_s^0$, and subtracting the contributions to
$F_2^{en}$ from those to $F_2^{ep}$, we obtain
\begin{eqnarray}
\lefteqn{(F_2^{ep} - F_2^{en})|_{soft}}\nonumber \\
 &=& \frac{I_{\pi}}{4f_{\pi}^2}[g_A^2(0)(F_2^{ep} - F_2^{en})(3<n> -1)
-16xg_A(0)(g_1^{ep} - g_1^{en})] ,
\end{eqnarray}
where $I_{\pi}$ is the phase space factor for the soft pion defined as
\begin{equation}
I_{\pi} = \int\frac{d^2\vec{k}^{\bot}dk^+}{(2\pi)^{3}2k^+},
\end{equation}
and $<n>$
is the sum of the nucleon and anti-nucleon multiplicity defined as\\
$<n>=<n>_p + <n>_n + <n>_{\bar{p}} + <n>_{\bar{n}}$.
\begin{figure}
    \epsfxsize = 15 cm   %or \epsfysize = 10 cm
    \centerline{\epsfbox{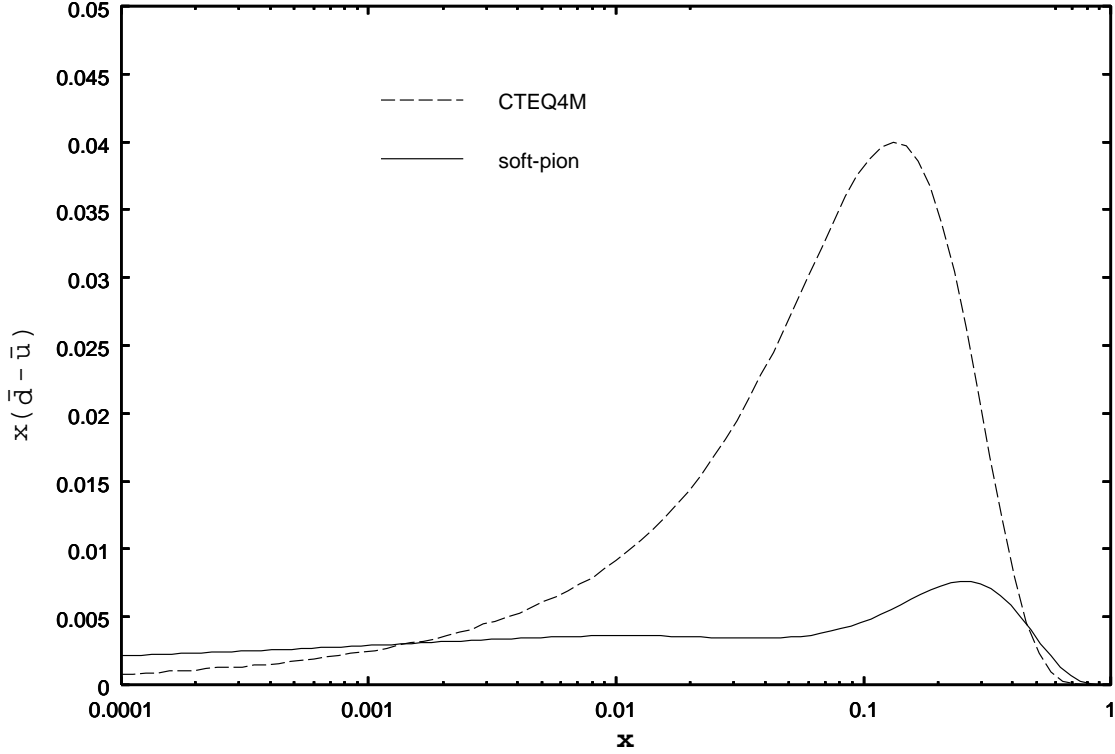}}
    \caption{Soft pion contribution to the flavor asymmetry of the light sea quarks}
    \label{fig:3}
\end{figure}
Note that the spin dependent term
is obtained by assuming the symmetric sea polariztion for simplicity. 
We can express $(F_2^{ep} - F_2^{en})|_{soft}$ as the asymmetry of the antiquark 
distribution as
\begin{equation}
(F_2^{ep} - F_2^{en})|_{soft} = -\frac{2}{3}x(\bar{d} - \bar{u})|_{soft}.
\end{equation}
To estimate the magnitude of this asymmetry, we approximate $F_2^{ep},
F_2^{en},g_1^{ep},g_1^{en}$ on the right-hand side of Eq.(5.1) by the
valence quarks distribution functions at $Q_0^2=4\; GeV^2$\cite{GS}.
As a multiplicity of the nucleon and antinucleon, we set
\begin{equation}
<n> = a\log_es +1 ,
\end{equation}
where $s=(p+q)^2$. The parameter $a$ is fixed as
0.2 in consideration for the proton and the anti-proton multiplicity
in the $e^+e^-$ annihilation such that $\frac{1}{2}a\log_es$ with 
$\sqrt{s}$ replaced by CM energy of that reaction agrees 
with it \cite{DELPHI}.
Following the experimental check of the charge asymmetry in the previous section,
the transverse momentum is restricted by the condition (1) and the Feynman scaling
variable is cut at $x_F=0.1$. By calculating the phase space factor under these
conditions and allowing a small change of the parameters to determine
the phase space, it is shown that we can expect the magnitude of the contribution to 
the Gottfried sum from the soft pion is about $-0.04 \sim -0.02$.  The main contribution
comes from the small $x$ region as is shown in Fig.3, where we give a plot of the same asymmetry
given by CTEQ4M\cite{CTEQ} for comparison.
\begin{figure}
    \epsfxsize = 15 cm   %or \epsfysize = 10 cm
    \centerline{\epsfbox{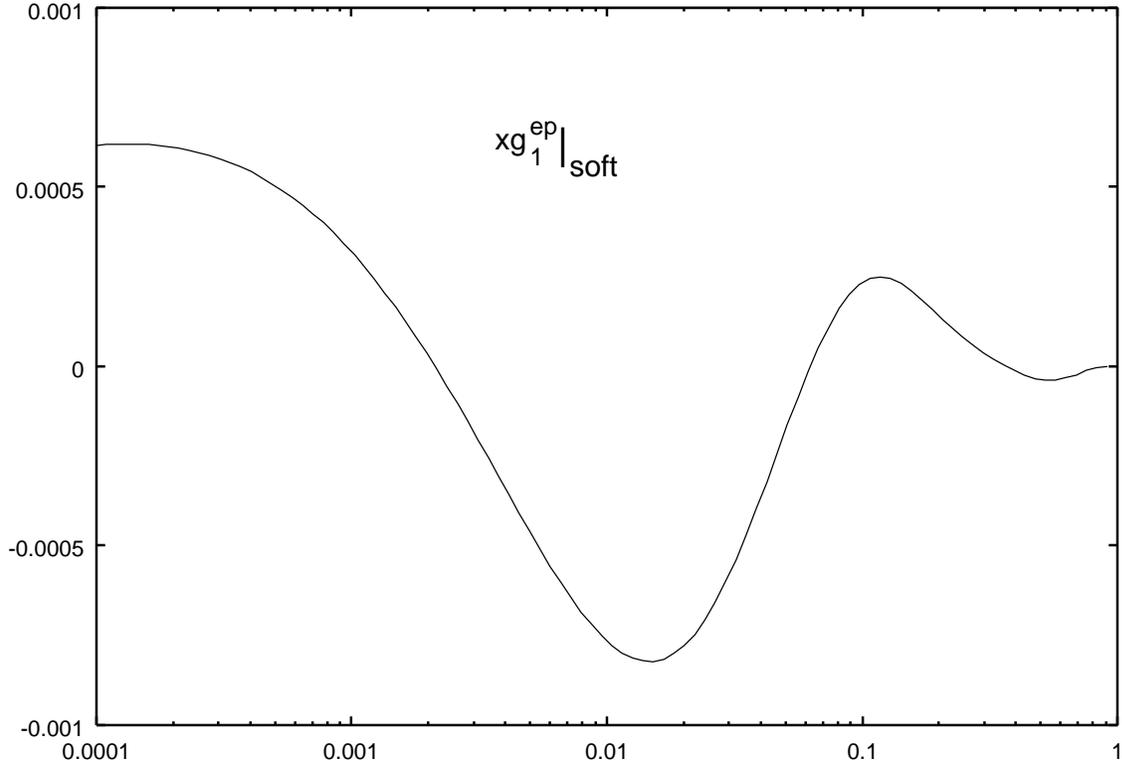}}
    \caption{Soft pion contribution to the $xg_1^{ep}$}
    \label{fig:4}
\end{figure}
\section{The soft pion contribution to the Ellis-Jaffe sum rule}
The soft pion contribution to the $g_1^{ep}$ is estimated under the same condition in the
previous section, and we obtain
\begin{eqnarray}
xg_1^{ep}|_{soft}
=&\frac{I_{\pi}}{4f_{\pi}^2}[-(xg_1^{\nu p}+xg_1^{\bar{\nu} p})+g_A^2(0)(xg_1^{ep}+2xg_1^{en})
+3g_A^2(0)<n>xg_1^{ep} \nonumber \\
&-3g_A(0)(F_2^{ep} - F_2^{en})-\frac{g_A(0)}{6}(F_2^{\nu p} - F_2^{\bar{\nu} p})].
\end{eqnarray}
Using the same quark distribution functions as in the previous section, the right-hand
side of Eq.(6.1) is estimated and the result is shown in Fig.4.
The contribution above $x=0.001$ is small, and we see that the contribution
from the small $x$ region may become large. However the effect of the soft pion above $x=0.0001$
to the Ellis-Jaffe sum rule is negligible. 
\section{Summary}
We have calculated the soft pion contribution to the Gottfried sum
and find that its magnitude is about $-0.04 \sim -0.02$. This magnitude is about
the same order required by the meson cloud model.
The same soft pion contribution to the Ellis-Jaffe sum rule in the region
above $x=0.0001$ is calculated, and find that
it is very small, but we find that the soft pion
contribution to $xg_1^p$ below $x=0.002$ becomes positive.

\end{document}